# SCALING ANALYSIS OF THE GALAXY DISTRIBUTION IN THE SSRS CATALOG

A. Campos

*Department of Physics, University of Durham, South Road, Durham DH1 3LE UK*

R. Domínguez-Tenreiro and G. Yepes

*Dpt. Física Teórica C-XI, Universidad Autónoma de Madrid, Cantoblanco 28049, Madrid SPAIN*

## ABSTRACT

A detailed analysis of the galaxy distribution in the Southern Sky Redshift Survey (SSRS) by means of the multifractal or scaling formalism is presented. It is shown that galaxies cluster in different ways according to their morphological type as well as their size. Ellipticals are more clustered than spirals, even at scales up to 15 $h^{-1}$ Mpc, whereas no clear segregation between early and late spirals is found. It is also shown that smaller galaxies distribute more homogeneously than larger galaxies.

*Subject headings:* galaxies: clustering – galaxies:fundamental parameters

## 1. INTRODUCTION

The understanding of the large scale distribution of the galaxies in the Universe is of key importance if one wants to test the different models proposed to explain its origin. That is not an easy task. On one hand, only a relatively small volume of universe has been mapped and that might well be not a fair sample drawn from an hypothetical galaxy distribution. Some studies on both the CfA1 (Huchra et al. 1983) and the SSRS (da Costa et al. 1988) galaxy catalogs (Davis et al. 1988 ; de Lapparent, Geller & Huchra 1988; Lachièze-Rey, da Costa & Maurogordato 1992) have raised the question as to whether these catalogs should be considered *fair* samples. On the other hand, we need better statistical estimators than the standard two-point correlation function, to account for the statistical properties of the distribution of galaxies. The multifractal formalism provides a deeper insight into the properties of point distributions than the two point correlation analysis. In fact, it has been shown (Martínez et al. 1990) that two distributions with the same two-point correlation function can considerably differ when they are described by the multifractal formalism.

As is well known, a complete statistical description will only be provided by the N-point correlation functions for any N (Peebles 1980). In fact, low- and high-order statistics are complementary, providing different insights into the properties of the galaxy distribution (Lachièze-Rey, da Costa & Maurogordato 1992). But unfortunately, it is difficult to measure correlation functions of order N>4 from current galaxy catalogs (Bonometto & Sharp 1980; Gaztañaga 1992). To circumvent this problem, a variety of new statistical descriptors have been used by several authors to study the galaxy clustering: topological analysis (Zel'dovich, Einasto & Shandarin 1982; Gott, Mellott & Dickinson 1986), Void Probability Function ( White 1979; Schaeffer 1984; Fry 1986; Maurogordato & Lachièze-Rey 1987; Fry et al. 1989; Vogeley, Geller & Huchra 1991; Lachièze-Rey, da Costa & Maurogordato 1992), count-in-cells (Efsthatiou et al. 1990) or the multifractal analysis (Jones et al. 1988; Martínez et al. 1990). In this work we have used the scaling (multifractal) analysis method, which provides specific statistical information of the point set from dense areas to rarefied ones, and at all range of scales.

It has been found (Einasto, Klypin & Saar 1986) that the correlation length $r_0$ in the CfA1 catalog increases with the sample volume. This effect, known as the *depth effect,* can be interpreted as a combined effect of sample location in the catalog and luminosity segregation (Maurogordato & Lachièze-Rey 1987, 1991; Xia, Deng & Zhou 1987; Hamilton 1988; Davis et al 1988). An alternative conclusion has been reached by Coleman & Pietronero (1992), who have argued that this effect is produced by the *fractal nature* of the Universe at any scale.

Any suitable cosmological model might explain the observed structures in the Universe. However there is a further complication due to the fact that the observed galaxy distribution may well not reflect the distribution of mass (Kaiser 1984). Any biased galaxy formation scenario will lead to segregation between galaxies (Dekel & Rees 1987) and therefore the search for segregation as a function of the galaxy properties is of great importance. In hierarchical clustering scenarios, halos are expected to be segregated as a function of their mass and density contrast (Mo, McGaugh & Bothun 1994). It is still not clear whether this segregation is better seen by analyzing the segregation of galaxies as a function of their morphology, size or luminosity, although it has been claimed that morphology has a stronger role in determining the clustering properties (Iovino et al. 1993; Cen & Ostriker 1993).

Besides the analysis of the galaxy distribution as a function of luminosity (Hamilton 1988; White, Tully & Davis 1988; Börner, Mo & Zhou 1989; Domínguez-Tenreiro & Martínez 1989; Salzer, Hanson & Gavazzi 1990; Santiago & da Costa 1990; Iovino et al. 1993; Domínguez-Tenreiro, Gómez-Flechoso & Martínez 1994) a major effort has been devoted to the search for segregation as a function of morphology and size. In this sense, it has been well established that ellipticals do cluster more strongly than spirals (Davis & Geller 1976; Dressler 1980; Postman & Geller 1984; Giovanelli, Haynes & Chincarini 1986). This has been recently confirmed using the CfA catalog as a database by Santiago & Strauss (1992) and Domínguez-Tenreiro, Gómez-Flechoso & Martínez (1994) who have shown that this segregation is detected up to large scales ($\geq 10h^{-1}$ Mpc) and in different density environments (Mo et al. 1992; Domínguez-Tenreiro et al. 1994). As the CfA and the SSRS catalogs are built up using different basis, i.e. the CfA is magnitude limited whereas the SSRS is diameter limited, it is of interest to check whether the morphological segregation is seen in the SSRS catalog as well. Mo et al. (1992), by means of the integral of the two-point correlation function, have detected morphological segregation at large scales in the SSRS catalog. The determination of the scales up to which there exists segregation as a function of morphology is an important clue to understanding the origin of the morphological differentiation (Oemler 1992; Salvador-Solé 1992). Therefore we find of great interest to check the existence of such large scale segregation by means of a different statistical approach.

More contradictory are the results concerning the segregation as a function of size. Dekel & Silk (1986) suggested that galaxies of different mass, (tentatively, of different size) should distribute in a different way, because they might have originated from peaks of different height of the density perturbation spectrum. Galaxies formed from $1\sigma$ peaks would be more homogeneously distributed. However Eder et al. (1989) as well as Thuan et al (1991) failed to find any segregation between dwarf and normal galaxies, although their samples were well biased toward blue, HI rich, objects (Campos, Moles & Masegosa 1993).

This paper is structured as follows: In section §2 we give a brief description of the statistical method we have used for our analysis. In §3 we describe the characteristics of the SSRS catalog and the samples we have worked with. Results of the analysis are presented in §4. Discussion and main conclusions can be found in §5. Throughout this work we have considered $H_0 = 100$ km s$^{-1}$ Mpc.

## 2. STATISTICAL DESCRIPTORS

The multifractal or scaling formalism was formulated as a generalization of the concept of fractals (Mandelbrot 1974, 1982). It was first applied to the study of strange attractors that appear in dynamical systems (Frish & Parisi 1985; Jensen et al 1985; Halsey et al 1986). The application of this formalism to the study of the large scale distribution of galaxies was first done by Jones et al. (1988). Here, we will briefly mention the basic ideas of the formalism and the algorithms used to deal with finite point sets. A more detailed description can be found in Martínez et al (1990), Yepes, Domínguez-Tenreiro & Couchman (1992) and Domínguez-Tenreiro, Gómez-Flechoso & Martínez (1994).

In a point set one can define a probability for each point of the set as:

$$p_i(r) = \frac{n_i(r)}{N}, \qquad (1)$$

where $n_i(r)$ is the number of points inside a sphere of radius $r$ centered at the point $i$ of the distribution and $N$ is the total number of points of the set.

Let us first consider the *correlation sum algorithm*, $Z(q,r)$. We define the following moments:

$$Z(q,r) \equiv \frac{1}{N_R} \sum_{i=1}^{N_R} p_i(r)^{q-1} \quad , \quad (q \neq 1) \qquad (2)$$

$$Z(1,r) \equiv \frac{1}{N_R} \sum_{i=1}^{N_R} p_i(r) \log p_i(r) \quad , \quad (q = 1) \qquad (3)$$

where each sum runs over $N_R$ randomly chosen points from the N total points in the set. In what follows we have taken $N_R$ equal to the total number of points in the sample.

The point set will have a multifractal structure if $Z(q,r)$ presents a scaling behavior, for all the $q$-moments, in a certain "scaling region" of $r$

$$Z(q,r) \propto r^{\tau_q} \qquad (4)$$

The function $\tau(q)$ is related to the so-called *generalized dimensions* (Hentschel & Procaccia 1983) by

$$D_q = \frac{\tau_q}{q-1} \qquad (5)$$

Depending on the $q$-moment, different areas of the set are weighted in different ways. For high positive $q$-moments, the sums in Eq (2) are dominated by the points of the higher density regions, while for high negative $q$-moments, the more rarefied areas of the set are statistically dominant. This is one of the main advantages of using this technique. We have an estimate of the statistical properties of a point set not only as a whole, but also for regions of different density.

It is not always possible to accurately compute $D_q$ for all values of $q$. For low $q$ values, the sums in (2) are dominated by points of poor statistics and the scaling behavior is broken. An alternative method has been proposed that is better suited to study low density regions. It is commonly called the *density reconstruction method* (Grassberger, Badii & Politi 1988) and is based in the computation of the moments

$$W(\tau,p) \equiv \frac{1}{N_R} \sum_{i=1}^{N_R} r_i(p)^{-\tau} \qquad (6)$$

where $r_i(p)$ is the radius of a sphere centered at the point $i$ with a probability $p = n/N$. Again, if there is a scaling region in $p$ where these functions scale like

$$W(\tau,p) \propto p^{1-q}, \qquad (7)$$

then the point distribution will have a multifractal structure. Eq (7) defines a $\tau(q)$ function that is related with the generalized dimensions by Eq (5).

We have used the $Z$ and $W$ functions to study the statistical properties of the galaxy distribution in the SSRS catalog. We note that $Z$ and $W$ are integral quantities, for instance, $Z(2,r) \propto J_3$. We have used Eq. (7) to estimate the generalized dimensions for $q < 2$ (termed $D^q$). Generalized dimensions for $q \geq 2$, $(D_q)$, have been estimated by means of Eq. (4).

If the samples do not occupy the same volume, the $Z$ and $W$ functions have to be re-scaled to the same volume, in order to compare the results for the different samples. It can be shown that (Domínguez-Tenreiro, Gómez-Flechoso & Martínez 1994) for large values of $r$, $Z(q,r) \propto (1/V)^{q-1}$ and for large probabilities, $p(r) \propto 1/V$ where $V$ is the total volume of the sample.

The more clustered the point distribution is, the less steep the $Z(q,r)$ and the $W(\tau,p)$ (for $D(q) > 3$) curves. For low $|\tau|$, more clustering implies steeper $W(\tau,p)$ curves.

We have applied the same boundary correction than Yepes, Domínguez-Tenreiro & Couchman (1992) applied to the study of CDM N-body simulations. A detailed description of the procedure can be found there.

In this work, we will show results corresponding to four different moments, $q = 2$ and $q = 5$ (for $Z$), and $\tau = -1$ and $\tau = -5$ (for $W$). These are representative of high density regions ($q = 5$), moderate density ($q = 2$ and $\tau = -1$) and rarefied areas ($\tau = -5$). The scaling behavior for the rest of the moments is qualitatively similar.

## 3. THE SSRS CATALOG

The Southern Sky Redshift Survey Catalog (SSRS; da Costa et al. 1991) consists of 2028 galaxies with major optical apparent diameters of $\sim 1.26''$ selected from the ESO/Uppsala Catalogs (Lauberts 1982). About 1600 galaxies have available redshifts, decreasing the completeness toward later types (beyond $S_c$). The objects are located in an area defined by $\alpha \in [5^h - 19^h]$, $\delta \leq -17°$ and $b \leq -30°$. We have corrected the galaxy redshifts for solar rotation, peculiar motion in the Local Group and for the infall of the Local Group toward the Virgo cluster with a infall velocity of $V_v = 440$ km s$^{-1}$ (Maurogordato & Lachièze-Rey 1987; 1991), The results shown in this paper are not sensitive to the particular value of $V_v$. A value of $q_0 = 0.05$ has been used to derive the galaxy distances.

To perform the analysis of the spatial distribution of galaxies in the SSRS catalog, we have extracted volume limited subsamples. Only those galaxies with absolute diameter larger than the threshold given by the sample's depth will be considered as belonging to the set. Most of the samples have a lower limit in redshift of 2000 km s$^{-1}$ because the presence of the Fornax cluster at about 1600 km s$^{-1}$ would distort the results of the analysis, as we will illustrate below. In what follows we specify the depth (in h$^{-1}$ Mpc) in the sample names. As an example, S40 corresponds to the sample limited by a minimum and maximum distance of 20 and 40 Mpc respectively whereas S040 corresponds to the sample limited only by a maximum redshift of 40 Mpc. In Table 1 we present the relevant parameters, such as minimum and maximum distance to the galaxies, number of galaxies, galaxy density, and limit in absolute diameter for all the galaxy samples that have been used in this work.

## 4. RESULTS

Using the algorithms described in §3, we have computed the $Z(r,q)$ and $W(p,\tau)$ functions for the samples described in Table 1. The two main goals of the present work were the analysis of the general properties of the distribution of galaxies, and the study of the segregation properties of the galaxies as a function of their morphological type and/or size.

In order to isolate the effects in the statistical descriptors produced by local features from those due to the intrinsic properties of the galaxy distribution, we have looked for systematic features in the $Z$ and $W$ functions corresponding to different samples of different depth. Only when a feature is *systematically* present, we could safely conclude that it is an intrinsic statistical property and not a local effect.

Table 1: Galaxy Samples

| Galaxy Sample | $d_{min}(h^{-1}Mpc)$ | $d_{max}(h^{-1}Mpc)$ | D (kpc) | Ngal | $\rho$ (gal/Mpc$^{-3}$) |
|---|---|---|---|---|---|
| S040 | 0 | 40 | 14.6 | 291 | $7.79 \times 10^{-3}$ |
| S050 | 0 | 50 | 18.3 | 306 | $4.20 \times 10^{-3}$ |
| S060 | 0 | 60 | 21.9 | 332 | $2.63 \times 10^{-3}$ |
| S020-D40 | 0 | 20 | 14.6 | 87 | $1.80 \times 10^{-2}$ |
| S020-D50 | 0 | 20 | 18.3 | 51 | $1.10 \times 10^{-2}$ |
| S020-D60 | 0 | 20 | 21.9 | 34 | $7.29 \times 10^{-3}$ |
| S40 | 20 | 40 | 14.6 | 204 | $6.24 \times 10^{-3}$ |
| S50 | 20 | 50 | 18.3 | 255 | $3.73 \times 10^{-3}$ |
| S60 | 20 | 60 | 21.9 | 298 | $2.45 \times 10^{-3}$ |
| S70 | 20 | 70 | 25.5 | 263 | $1.34 \times 10^{-3}$ |
| S80 | 20 | 80 | 29.2 | 211 | $7.17 \times 10^{-4}$ |
| S50-D70 | 20 | 50 | 25.5 | 83 | $1.21 \times 10^{-3}$ |
| S60-D70 | 20 | 60 | 25.5 | 175 | $1.44 \times 10^{-3}$ |
| S60-D80 | 20 | 60 | 29.2 | 100 | $8.24 \times 10^{-4}$ |
| S70-D80 | 20 | 70 | 29.2 | 147 | $7.52 \times 10^{-4}$ |
| S40c | 20 | 31 | 14.6 | 104 | $8.18 \times 10^{-3}$ |
| S40d | 31 | 40 | 14.6 | 100 | $5.0 \times 10^{-3}$ |
| S50c | 20 | 42 | 18.3 | 128 | $3.32 \times 10^{-3}$ |
| S50d | 42 | 50 | 18.3 | 127 | $4.27 \times 10^{-3}$ |
| S60c | 20 | 50 | 21.9 | 150 | $2.13 \times 10^{-3}$ |
| S60d | 50 | 60 | 21.9 | 148 | $2.91 \times 10^{-3}$ |
| S70c | 20 | 55 | 25.5 | 133 | $1.4 \times 10^{-3}$ |
| S70d | 55 | 70 | 25.5 | 130 | $1.29 \times 10^{-3}$ |
| S80c | 20 | 61 | 29.2 | 105 | $8.22 \times 10^{-4}$ |
| S80d | 61 | 80 | 29.2 | 106 | $6.37 \times 10^{-4}$ |
| S40E | 20 | 40 | 14.6 | 40 | $1.38 \times 10^{-3}$ |
| S40Se | 20 | 40 | 14.6 | 62 | $1.89 \times 10^{-3}$ |
| S40Sl | 20 | 40 | 14.6 | 89 | $2.72 \times 10^{-3}$ |
| S50E | 20 | 50 | 18.3 | 57 | $8.34 \times 10^{-4}$ |
| S50Se | 20 | 50 | 18.3 | 86 | $1.28 \times 10^{-3}$ |
| S50Sl | 20 | 50 | 18.3 | 104 | $1.52 \times 10^{-3}$ |
| S60E | 20 | 60 | 21.9 | 70 | $5.77 \times 10^{-4}$ |
| S60Se | 20 | 60 | 21.9 | 109 | $8.98 \times 10^{-4}$ |
| S60Sl | 20 | 60 | 21.9 | 113 | $9.31 \times 10^{-4}$ |
| S70E | 20 | 70 | 25.5 | 58 | $2.91 \times 10^{-4}$ |
| S70Se | 20 | 70 | 25.5 | 106 | $5.43 \times 10^{-4}$ |
| S70Sl | 20 | 70 | 25.5 | 96 | $4.92 \times 10^{-4}$ |
| S80E | 20 | 80 | 29.2 | 40 | $1.32 \times 10^{-4}$ |
| S80Se | 20 | 80 | 29.2 | 84 | $2.85 \times 10^{-4}$ |
| S80Sl | 20 | 80 | 29.2 | 85 | $2.89 \times 10^{-4}$ |
| S40s | 20 | 40 | 14.6-18.2 | 104 | $3.18 \times 10^{-3}$ |
| S40l | 20 | 40 | 18.2-$\infty$ | 100 | $3.06 \times 10^{-3}$ |
| S50s | 20 | 50 | 18.3-22.7 | 127 | $1.86 \times 10^{-3}$ |
| S50l | 20 | 50 | 22.7-$\infty$ | 128 | $1.86 \times 10^{-3}$ |
| S60s | 20 | 60 | 21.9-26.7 | 149 | $1.23 \times 10^{-3}$ |
| S60l | 20 | 60 | 26.7-$\infty$ | 149 | $1.23 \times 10^{-3}$ |
| S70s | 20 | 70 | 25.5-30.2 | 132 | $6.75 \times 10^{-4}$ |
| S70l | 20 | 70 | 30.2-$\infty$ | 131 | $6.75 \times 10^{-4}$ |
| S80s | 20 | 80 | 29.2-34.2 | 106 | $3.60 \times 10^{-4}$ |
| S80l | 20 | 80 | 34.2-$\infty$ | 105 | $3.60 \times 10^{-4}$ |

In Figures 1 and 2 we show the $Z$, (for $q = 2$ and $q = 5$), and W, (for $\tau = -1$ and $\tau = -5$) functions, respectively, corresponding to the samples S40, S60 and S80. From these Figures it can be seen that there exists an scaling behavior in the $Z$ and $W$ functions. These results are in agreement with other works using the CfA redshift catalog (Domínguez-Tenreiro, Gómez-Flechoso & Martínez 1994). This scaling behavior is somehow destroyed at small distances (or, equivalently, small probabilities) due to discreteness effects caused by the small spatial density of points in the samples. This effect is much more important for the less dense samples, as can be seen in Figure 1.

### 4.1. Error Analysis

The statistical errors in the calculation of the $Z$ and $W$ functions were estimated using the bootstrapping resampling technique (Bradley 1982; Ling, Frenk & Barrow 1986). In the plots shown throughout this work error bars correspond to $2\sigma$ deviation calculated by means of 400 bootstrap realizations of each sample. The magnitude of the error bars depends on the scale considered. The errors decrease for larger scales and higher probabilities. On average we have found that at scales of $r \sim 1$ h$^{-1}$ Mpc the statistical uncertainty is of about 11% for $q = 2$ and can be as high as a factor of three for $q = 5$. For $r \geq 5$ h$^{-1}$ Mpc errors are less than 2% for $q = 2$ and 30% for $q = 5$. For the $W$ functions the errors are less than 1% for $\tau = -1$ and 4% for $\tau = -5$ for the range of probabilities we have considered.

To estimate the stability of the algorithms against variations in the number of points, we extracted about two hundred subsamples with half the number of galaxies randomly selected from the corresponding parent sample. We compared the results of the error analysis using the bootstrapping technique with those obtained by this dilution method. We found that the magnitude of the errors is similar in both cases. An example of this is shown in §4.5.

### 4.2. Effect of the Fornax Cluster

The scaling analysis of N-body simulations (Valdarnini, Borgani & Provenzale 1992; Yepes, Domínguez-Tenreiro & Couchman 1992) has shown that gravitational clustering evolution yields $D_\infty \to 1$. This can be understood as a reflection of the $\rho(r) \propto r^{-2}$ behavior in the denser areas of the point distribution, which is naturally reached in nonlinear gravitational evolution. The effect of peculiar velocities distorts severely the point distribution as seen in the redshift space, giving rise to an increase in the $D_\infty$ value (Yepes et al. 1994). This result can be used in the analysis of galaxy catalogs as an indicator of the contamination by peculiar velocities of the sample considered.

The nearby Fornax cluster, which is located at a distance of about 1600 km s$^{-1}$, is present in the SSRS catalog. To remove this cluster from the catalog we have considered only those galaxies with velocities greater than 2000 Km s$^{-1}$. To select this limit we made sure that the results obtained were very similar to those obtained by considering higher limits (up to 3000 km s$^{-1}$). Therefore the effect(s) that this cluster could introduce are removed by considering only galaxies with velocities greater than 2000 Km s$^{-1}$. By using a minimum limit, we also remove nearby galaxies from the samples whose low redshifts are expected to be dominated by peculiar motions. The comparison of samples with and without this limit can be then used to investigate the effects of the peculiar velocities on the statistical descriptors.

In Figure 3 we show the $D_q$ and $D^q$ dimensions for the scaling regions of $1h^{-1}$ Mpc $\lesssim r \lesssim 8h^{-1}$ Mpc and $0.03 \lesssim p \lesssim 0.4$ respectively, for the samples S040 / S40, S050 / S50 and S060 / S60. It can be clearly seen that $D_\infty$ takes higher values in those samples in which the Fornax cluster is present, and are rather independent of the sample depth. The values of $D_\infty$ are slightly different in those samples without Fornax, decreasing with the depth of the samples. In Table 1 we can see that the galaxy density corresponding to the volume 00-20 for the S020-D40 (i.e. galaxies in the 00-20 volume but with the limiting size corresponding to 40 $h^{-1}$ Mpc), S020-D50 and S020-D60 samples is $\sim$ 3 times higher than the galaxy density in the volumes 20-40 (S40), 20-50 (S50) and 20-60 (S60). Therefore those samples which contain Fornax are statistically dominated by this cluster.

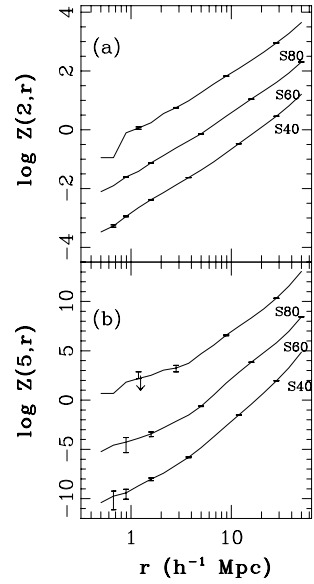

Fig. 1.— $Z(2,r)$ (a) and $Z(5,r)$ (b) for the samples S40, S60 and S80. $Z$ values have been shifted in order to show the three data sets together.

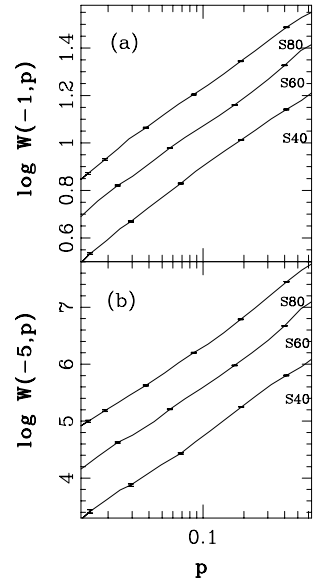

Fig. 2.— $W(-1,p)$ (a) and $W(-5,p)$ (b) for the samples S40, S60 and S80.

It is interesting to note that the correlation dimension ($D_2$) does not vary when Fornax is removed. This is an important result because the correlation dimension is related to the slope ($\gamma$) of the two-point correlation function ($\gamma \approx 3 - D_2$). To illustrate this, we plot in Figure 4 the $Z(2, r)$ corresponding to the samples S040 / S40, S050 / S50 and S060 / S60. We see that the results (for $q = 2$) are independent of the presence of Fornax. This is due to the fact that Fornax is a high density area and thus will dominate the results for the higher order moments (i.e. $q > 2$) in the S040, S050 and S060 samples. It can be seen in Eq. (2) that for $q = 2$, the areas of moderate density are statistically dominant in the computation of $Z$. Therefore, the effects of Fornax are not very important, as it is shown in the Figure. In what follows we will only consider galaxies with $v > 2000$ km s$^{-1}$.

### 4.3. Catalog Inhomogeneities

To study the effects in the statistical descriptors of location effects in the catalog, we have divided each sample into two disjoint subsamples (i.e. each sample is split into a close one and a distant one) and have compared the $Z$ and $W$ functions obtained from them. In Table 1 we present the parameters corresponding to each subsample (in the Table, the $c$ and $d$ in the sample names are for the close and distant samples respectively). To avoid problems related to the stability of the $Z$ and $W$ functions against differences in the number of points in the sets (see Domínguez-Tenreiro, Gómez-Flechoso & Martínez 1994) we have split each sample in such a way that the two corresponding subsamples have the same number of galaxies. This, however, makes the subsamples to comprise different volumes. Therefore, it was necessary to re-scale the $Z$ and $W$ functions to the same volume. As it can be seen in Table 1, the two subsamples have the same limit in diameter for all cases. Therefore, the differences in the statistical descriptors must be due to real inhomogeneities in the catalog, because the disjoint subsamples are built up by the same kind of galaxies. In Figures 5 and 6 we show $Z$ and $W$ respectively. The differences found are not systematic, contrary to those that will be discussed below. This is what is expected if the differences are the product of local inhomogeneities between differents regions of the catalog. As an example, let us notice that in the volume 20-40 and for regions of intermediate to high density (i.e. $q = 5$) the galaxies in the "distant" subsample are more clustered than those in the "close" subsample, occurring just the opposite in the 20-60 volume. Results for the volumes 20-50 and 20-70 (not shown in the Figures) are similar in the sense of absence of systematic features.

These differences between the disjoint subsamples, represent an estimate of the variance in the galaxy distribution in the catalog, which is clearly larger than the pure statistical errors of the algorithms calculated by the bootstrapping technique.

### 4.4. Depth effect

Einasto, Klypin & Saar (1986) noticed that the correlation length $r_0$ in the CfA catalog increases with the sample volume. This is the so-called *depth effect.* Other authors have also shown the existence of such effect using low order statistical descriptors, in optically selected galaxy catalogs (Davis et al. 1988). The existence of a *depth effect* has been explained as a combination of factors: large positive density fluctuations in the foreground of the optically selected catalogs, and/or the existence of luminosity (or size) segregation. The distribution of IRAS galaxies, on the contrary, seems to be fairly independent on the scale (Davis te al. 1988).

In Figure 7 we show the $Z$ functions for the S40, S60 and S80 samples, re-scaled to the same volume. We can clearly see that the *depth effect* is not only present in the low-order moments ($Z = 2$) but also in the high-order ones ($Z = 5$). Moreover, the effect is much more clearly seen in the latter.

This effect could be caused by size segregation, i.e. larger galaxies (those seen in deeper samples) being more strongly clustered than smaller ones. It could also be due to contamination by peculiar velocities, which is more important for the closer samples. To check this, we have considered the samples sets S50-D70 / S60-D70 / S70 and S60-D80 / S70-D80 / S80, where D70 and D80 in the sample names mean that the limit in diameter corresponds to a depth of 70 and 80 $h^{-1}$ Mpc respectively (see Table 1). The results for the $Z$ and $W$ functions are shown in Figures 8 and 9 respectively. It can be seen in the Figures that there is no *depth effect* at all when we compare samples of different depths but with galaxies in the same size ranges, and the rather small differences found are not systematic. The fact that the *depth effect* vanishes when galaxies in the same range of sizes are considered means that the differences seen in Figure 7 are actually a by-product of size segregation.

### 4.5. Morphological Segregation

We have studied if galaxies of different morphological types share the same spatial distribution, or if they are distributed in space in different ways. For this analysis of the spatial distribution as a function of morphology, we have divided the samples into three subsamples, according to the galaxy's morphological code (i.e. *T*-type): Ellipticals and Lenticulars (E) with T $\leq 0$, Early Spirals (Se) with $1 \leq$ T $\leq 4$ and Late Spirals (Sl) with $5 \leq$ T $\leq 8$. We have not considered irregulars because we are dealing with large galaxies and they are present in the subsamples in a very low percentage. Galaxies without morphological classification (i.e. dwarf galaxies, interacting pairs, etc) have also been excluded. As it can be seen in Table 2, the percentage of excluded galaxies is less than about 5% in all samples.

In Figures 10 and 12 we show $Z$ and $W$ corresponding to the three morphological subsamples in the volumes 20-40, 20-50 and 20-60. (The analysis was also performed for the S70 and S80 samples, finding consistent results. Although the number density of galaxies, ellipticals in particular, is now quite low as can be seen in Table 1). For this analysis we could not force the number of galaxies in each subsample to be the same, as we did previously when we studied the catalog inhomogeneities. To check the stability of the algorithms against differences in the number of points, we extracted 200 different subsamples randomly selecting, from the spiral samples (early + late), which are always larger, the same number of galaxies than in the corresponding (i.e. same volume) elliptical samples.

Table 2: T-types versus Sizes

| Galaxy Sample | %(Ellipticals) | %(Spirals) |
|---|---|---|
| S40s | 34 | 67 |
| S40l | 14 | 84 |
| S50s | 16 | 80 |
| S50l | 25 | 76 |
| S60s | 23 | 71 |
| S60l | 21 | 75 |
| S70s | 23 | 72 |
| S70l | 20 | 76 |
| S80s | 23 | 70 |
| S80l | 13 | 79 |

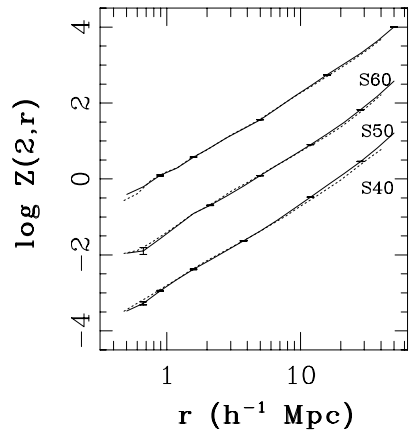

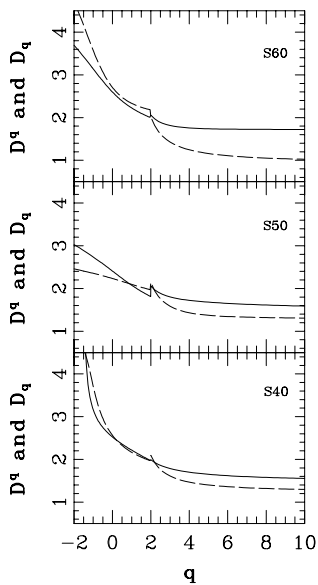

Fig. 3.— $D^q$ and $D_q$ versus $q$ calculated for the volumes S40, S50 and S60, (dashed lines), and S040, S050 and S060, (solid lines). The scaling regions considered are 1 $h^{-1}$ Mpc $\leq r \leq 8$ $h^{-1}$ Mpc ($q \geq 2$) and $0.03 \leq p \leq 0.4$ ($q < 2$).

Fig. 4.— $Z(2, r)$ for the volumes S40, S50 and S60 (dashed lines) and S040, S050 and S060 (solid lines). The $Z$ values have been re-scaled to the same volume for samples of the same depth.

We then computed $Z$ for these sets, and calculated the mean value. The results are presented in Figure 11. The error bars in the spiral subsamples are 2 standard deviation computed by means of this *dilution* method, while error bars in ellipticals are $2\sigma$ estimated with the bootstrapping technique.

In regions of medium to high density, described by $Z(q, r)$, we see that there is a clear difference between ellipticals and spirals, being the former more clustered than the latter. For $\tau = -1$, the same result is found. For all the samples analyzed, we found that, systematically, $Z_{ellip} > Z_{sp}$ and $W(-1, p)_{ellip} < W(-1, p)_{sp}$, at least, at a $3\sigma$ significance level. Therefore it seems safe to conclude that elliptical galaxies seem to be more clustered than spirals, *in scales up to 15 $h^{-1}$ Mpc and even larger*. It is interesting to remark that deeper samples contain larger galaxies. This means that, regardless of their sizes, ellipticals are more strongly clustered than spirals.

Both the $Z$ and $W$ functions are integral quantities, and so the clustering at small scales could be propagated in the integral to the larger scales. That is, the segregations seen at large scales could be produced by the rich clusters in the catalog. However, it is known that there are no rich galaxy clusters in the SSRS catalog, such as the Coma cluster in the CfA. The richest one is Fornax, which is not a very massive cluster ($R = 0$ Abell cluster).

In the study of morphological segregation in the CfA cata-log by means of the $Z$ and $W$ functions (Domínguez-Tenreiro et al. 1994), it was checked that the results were not altered when Coma galaxies were excluded in the sample. The same conclusion was drawn by Mo et al (1992). In this work the morphological segregation was studied by means of the integral of the two point correlation function (which is equivalent to $Z(2,r)$). To check the robustness of their results, they repeated the analysis removing the rich clusters in the catalogs, obtaining similar results. As we already showed in Figure 3 (see §2.1), the same results for $Z(2,r)$ are obtained when Fornax is excluded from the sample. Moreover, as it can be seen in Figures 10-12, the segregation between spirals and ellipticals is not only seen for high order moments, where the role of the high density regions such as clusters is more important, but also in low order moments ($q = 2$ and $\tau = -1$) where more weight is given to regions of moderate density.

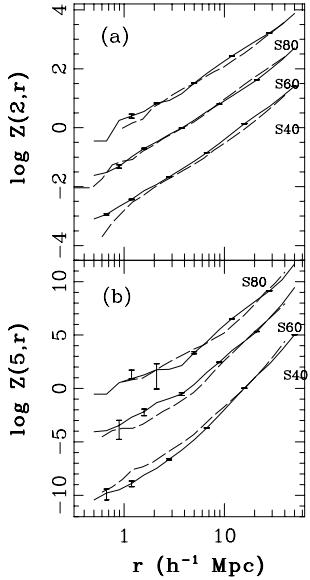

Fig. 5.— $Z(2,r)$ (a) and $Z(5,r)$ (b) for the following sets: S40c, S60c and S80c (solid lines) and S40d, S60d and S80d (dashed lines).

Results for early and late spiral subsamples are not so clear. In the volume 20-40 and for the high density regions we find that late types seem to be more strongly clustered than early types, contrary to what would be expected. For the rest of the deeper volumes, early and late type spirals share the same kind of clustering. In the rarefied areas early and late types present the same behavior except for the volume 20-80, not shown in the Figure, where again late types are more clustered than early ones.

### 4.6. Size Segregation

In Figure 7 we compare results for the S40/S60/S80 samples normalized to the same volume. It can be seen that galaxies in the larger volumes appear to be more strongly clustered than galaxies in the smaller ones. This could be understood in terms of size segregation, because the limiting size of the samples is different (see Table 1) and we checked (see §2.3) that there is no *depth effect* when galaxies in the same size ranges are considered.

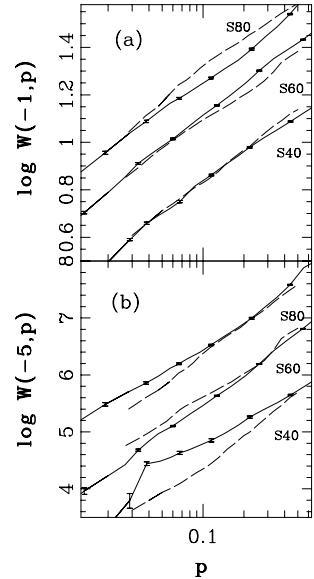

Fig. 6.— $W(-1,p)$ (a) and $W(-5,p)$ (b). Data samples are as in Fig 4.

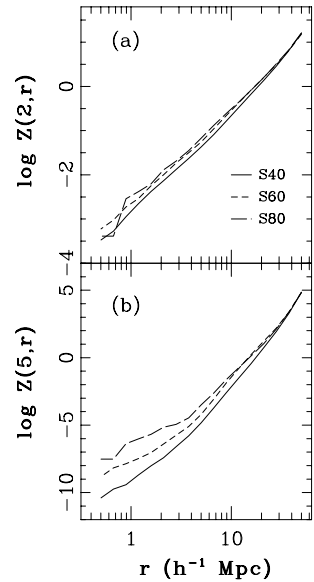

Fig. 7.— Same than in Fig. 1, but the samples have been rescaled to the same volume to illustrate the existence of a *depth effect* in both the low- and high-order statistical descriptors.

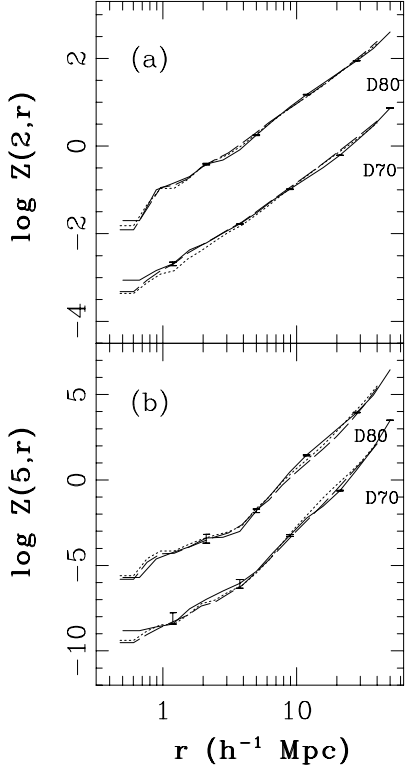

Fig. 8.— $Z(2,r)$ (a) and $Z(5,r)$ (b) for the following samples (see Table 1): Labeled with D70: S50-D70 (solid line), S60-D70 (dashed line) and S70 (dot-dashed line). Labeled with D80: S60-D80 (solid line), S70-D80 (dashed line) and S80 (dot-dashed line).

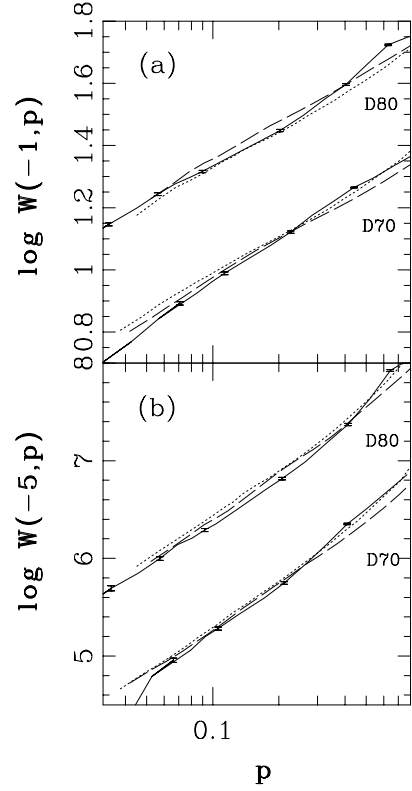

Fig. 9.— $W(-1,p)$ (a) and $W(-5,p)$ (b). Data samples are as in Fig 8.

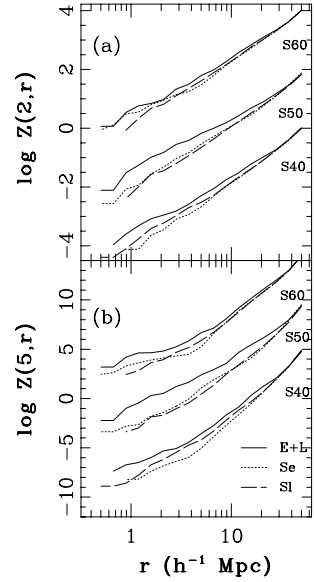

Fig. 10.— $Z(2,r)$ (a) and $Z(5,r)$ (b) for the S40, S50 and S60 samples and for 3 different galaxy types: elliptical galaxies (solid lines), early spirals (dot-dashed lines) and late spirals (dashed lines).

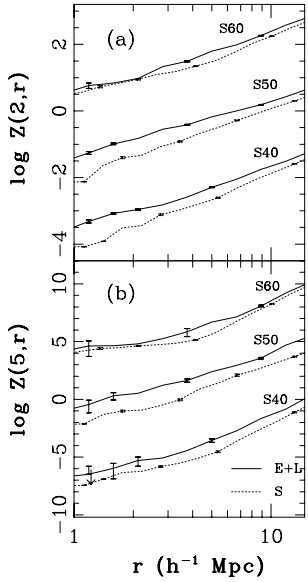

Fig. 11.— Same as Fig 10 but now for two galaxy types: ellipticals (solid lines) and spirals (early + late; dashed line).

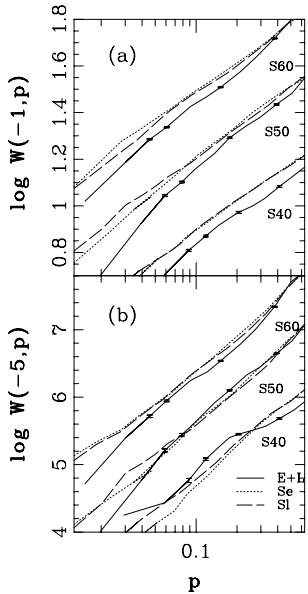

Fig. 12.— $W(-1,p)$ (a) and $W(-5,p)$ (b). Data samples are as in Figure 10.

To test this hypothesis, we divided the samples into two subsamples, one corresponding to *small* galaxies and the other to *large* galaxies (named $s$ and $l$ in the sample names respectively, see Table 1). For the 5 volumes studied, the limits in diameter of the subsamples are different, because we forced the number of galaxies in the two subsamples corresponding to each volume to be similar as we did in §4.2. Results for the $Z$ and $W$ functions are shown in Figures 13 and 14 respectively. Error bars in the Figures are $2\sigma$ deviation estimated with the bootstrapping technique. Size segregation is clearly seen in the S50 and S60 samples, as well as in the S80 for $\tau = -1$ and -5 (the behavior for the S70 volume, not shown in the Figure, is similar to the S80 one). For the S40 some small segregation is detected at smaller scales for $q = 2$ and $\tau = -1$.

## 5. DISCUSSION AND CONCLUSIONS

Let us first briefly summarize the results shown in the previous sections.

1. We have used the $Z$ and $W$ functions to study the clustering properties of galaxies in the SSRS catalog. The main interest of using these functions for the statistical study of segregation properties of galaxies is that regions of different density can be analyzed separately. A total of 44 different diameter-limited samples have been extracted and analyzed from the catalog.

2. We removed from the samples all those galaxies with velocity less than 2000 km s$^{-1}$. The reason is the presence of the nearby Fornax cluster at 16 h$^{-1}$ Mpc. As we are working in redshift space, the distorsion of real structures by peculiar motions produce undesirable effects in the $Z$ and $W$ functions. This is particularly true for those high density regions such as rich clusters, where the peculiar velocity field is stronger.

3. There exists a *depth effect* in the catalog, that occurs in all the $Z$ moments, being larger for the high order ones. We checked that the *depth effect* vanishes when galaxies in the same size range are considered. Therefore, the effect is more likely produced by the existence of size segregation rather than by variations in the density on the scales of the sample sizes.

4. We have found that elliptical galaxies are more clustered than spirals, even at scales larger than galaxy clusters. This is true not only for high density regions but also for regions of medium to low density. We also looked for differences between early and late spirals. The main result was that $Z$ and $W$ for both types of galaxies present the same behavior (except in two volumes), in high density regions.

5. Finally, we have also found that large galaxies seem to be slightly more clustered than small galaxies, although this is not outstanding in all the analyzed volumes. The detected segregation is nevertheless not very big.

We have shown that ellipticals are more clustered than spirals, not only in high density regions, such as galaxy clusters, but also in more rarefied areas. Probably, the most interesting result comes from the fact that segregation is seen at scales up to 15 h$^{-1}$ Mpc and even larger (i.e. larger than galaxy clusters). This result is consistent with what has been found in the CfA catalog (Santiago & Strauss 1992; Mo et al 1992; Domínguez-Tenreiro, Gómez-Flechoso & Martínez 1993), and in the SSRS catalog using different statistical techniques (Mo et al. 1992).

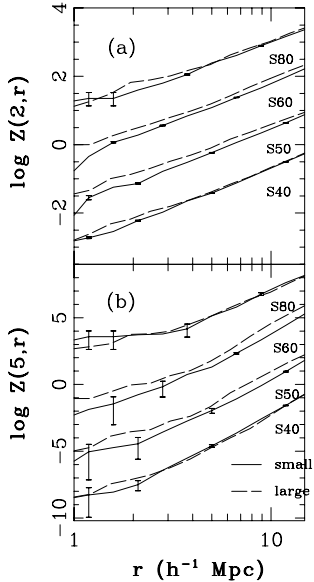
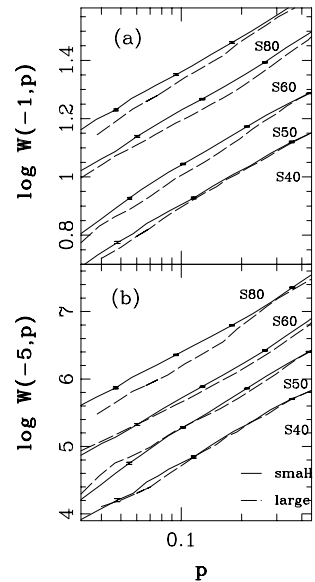

Fig. 13.— $Z(2,r)$ (a) and $Z(5,r)$ (b) for the S40, S50 S60 and S80 samples, and for two different galaxy sizes: small galaxies (solid lines) and large galaxies (dashed lines).

Fig. 14.— $W(-1,p)$ (a) and $W(-5,p)$ (b). Data samples are in Fig. 11.

It is interesting to remark that, by means of the multifractal formalism, morphological segregation up to 10-15 $h^{-1}$ Mpc has been found in both the CfA and the SSRS catalog. At first sight, the CfA and SSRS show different structures. As an example, in the SSRS rich clusters are not easy recognizable and it contains more galaxies of later types than the CfA (da Costa et al. 1988). Also there are less SSRS galaxies in groups in respect to the CfA (Maia & da Costa 1990) In principle, some differences could be due to the fact that the CfA catalog is magnitude limited, whereas the SSRS is diameter limited (Bardelli et al. 1991). In any case there are evidences that these catalogs are not fair samples because the observed structures have sizes comparable to the survey depths (deLapparent, Geller & Huchra 1988). Therefore it is of major interest to check whether features such as large-scale morphological segregation are observed in both catalogs, as they are statistically independent samples.

To understand the taxonomical and structural differences between spirals and ellipticals, many authors have speculated with the hypothesis that ellipticals are the results of galaxy mergers. This idea is supported by the well known strong clustering of ellipticals, which are usually located in high density regions, avoiding isolated areas. We have however shown that the differences in clustering are present at all range of scales, up to 15 $h^{-1}$ Mpc. Galaxy interactions or mergers could not easily explain the observed differences in the clustering properties at these large scales. On the contrary, this result could support the "ab initio" scenario for the origin of galaxy morphology. We then would be detecting the difference in clustering between the evolved distribution of density peaks that produced ellipticals and the evolved distribution of density peaks from which spirals formed (Cen & Ostriker 1993).

Small galaxies seem to be less clustered than larger galaxies, although the results in this case are much weaker than for the morphological segregation. The existence of size segregation is very probably the mechanism responsible for the *depth effect* observed in the SSRS catalog. Size segregation is in some sense equivalent to mass segregation In fact, mass segregation could be produced by the effect of biasing in the galaxy formation process (Kaiser 1984) and/or as a by-product of the violent relaxation phase in high density regions. In a Cold Dark Matter scenario it has been suggested that the correlation length is a strong function of the circular velocity of the galaxy (White et al. 1987) and so of the luminosity (or mass) through the Tully-Fisher relation. Let us notice, however, that *isophotal* size is related to mass as long as galaxies have all similar surface brightness. This is partially true for the SSRS catalog, as most of the galaxies are *high* surface brightness galaxies, although we might bear in mind that the correlation size - mass is relatively rough.

All the analysis that is presented in this work have been done in redshift space, i.e. the distances to the galaxies have been computed without doing any correction for peculiar velocities (except peculiar motion of the Local Group) We should question how this fact affects the results found about segregation (both in size and type). From the analysis of N-body simulations in real and redshift space (Yepes et al 1994) it is known that the $Z$ functions are very sensitive to the peculiar velocity field. The $Z$ functions describe a less clustered distribution in redshift space compared with real space. This difference in clustering is proportional to the peculiar velocity field of the particle set. In the SSRS catalog, large galaxies as well as elliptical galaxies are more clustered than small and spiral ones. It would be expected that large galaxies and ellipticals were more affected by peculiar motions because they will be mainly located in clusters. Therefore, the difference in clustering between large/small or ellipticals/spirals should be larger in real space than in redshift space, at least, up to cluster scales. For larger scales (10-15 $h^{-1}$ Mpc) where we do still see some degree of segregation, it is expected that the galaxy distribution will not be very strongly affected by peculiar motions.

Finally, we would like to point out that both classes of segregation (i.e. by type and by size) are separate effects. Spirals and ellipticals share approximately the same mass range and thus should have, by effect of biasing, similar clustering properties. We can see in Table 2 that they are present in similar percentages in all volumes, (i.e. for all the size range). The fact that there is no correlation between size and type ensures also that we can study separately clustering as a function of size from clustering as a function of morphological type.

AC wishes to thank LAEFF (INTA, Spain) for allowing her to use its computing facilities. This work has been partially supported by the DGICyT (Spain) under project numbers PB90-0182 and AEN90-0272.